\documentclass[10pt,twocolumn]{IEEEtran}
\usepackage{graphicx}          
\usepackage{amsmath}
\usepackage{tabularx}
\usepackage{amsfonts}
\usepackage{url}
\usepackage{verbatim}
\usepackage{times,epsfig}
\usepackage{makecell}
\usepackage{psfrag}
\usepackage{subfigure}
\usepackage{stfloats}
\usepackage{setspace}
\usepackage{amssymb}
\usepackage{multirow}
\usepackage{colortbl}
\usepackage[justification=centering]{caption}
\usepackage{fancyhdr}
\usepackage[table,xcdraw]{xcolor}

\begin{document}
	
	\pagenumbering{arabic}
	
	%\title{A Secure and Reliable Transfer Learning Framework for Emerging 6G Services and	Applications}
	% Quantum Key Distribution Service Provision
	\title{Quantum-Secured Space-Air-Ground Integrated Networks: Concept, Framework, and Case Study}
	\author{Minrui Xu, Dusit Niyato, Zehui Xiong, Jiawen Kang, Xianbin Cao, Xuemin Sherman Shen, and Chunyan Miao
	\thanks{Minrui~Xu, Dusit~Niyato, and Chunyan~Miao are with the School of Computer Science and Engineering, Nanyang Technological University, Singapore (e-mail: minrui001@e.ntu.edu.sg; dniyato@ntu.edu.sg; ascymiao@ntu.edu.sg).}
	\thanks{Zehui~Xiong is with the Pillar of Information Systems Technology and Design, Singapore University of Technology and Design, Singapore 487372, Singapore (e-mail: zehui\_xiong@sutd.edu.sg).}
	\thanks{Jiawen~Kang is with the School of Automation, Guangdong University of Technology, China (e-mail: kavinkang@gdut.edu.cn).}
	\thanks{Xianbin Cao is with the School of Electronic and Information Engineering, Beihang University, Beijing 100191, China. (e-mail: xbcao@buaa.edu.cn).}
	\thanks{Xuemin~Sherman~Shen is with the Department of Electrical and Computer Engineering, University of Waterloo, Waterloo, ON, Canada, N2L 3G1 (e-mail: sshen@uwaterloo.ca).}
	}
	\maketitle
	\pagestyle{headings}

	\begin{abstract}
		In the upcoming 6G era, existing terrestrial networks have evolved toward space-air-ground integrated networks (SAGIN), providing ultra-high data rates, seamless network coverage, and ubiquitous intelligence for communications of applications and services. However, conventional communications in SAGIN still face data confidentiality issues. Fortunately, the concept of Quantum Key Distribution (QKD) over SAGIN is able to provide information-theoretic security for secure communications in SAGIN with quantum cryptography. Therefore, in this paper, we propose the quantum-secured SAGIN which is feasible to achieve proven secure communications using quantum mechanics to protect data channels between space, air, and ground nodes. Moreover, we propose a universal QKD service provisioning framework to minimize the cost of QKD services under the uncertainty and dynamics of communications in quantum-secured SAGIN. In this framework, fiber-based QKD services are deployed in passive optical networks with the advantages of low loss and high stability. Moreover, the widely covered and flexible satellite- and UAV-based QKD services are provisioned as a supplement during the real-time data transmission phase. Finally, to examine the effectiveness of the proposed concept and framework, a case study of quantum-secured SAGIN in the Metaverse is conducted where uncertain and dynamic factors of the secure communications in Metaverse applications are effectively resolved in the proposed framework.
		
		% I think here we can describe some performance evaluations and briefly mention why the proposed one is effective.
		
		% we propose an efficient QKD service provisioning framework to support green and secure communication in SAGIN. Specifically, we propose a collaborative optical fiber, satellite and UAV-based QKD service provisioning system to meet the heterogeneous and uncertain security requirements in SAGIN. In terrestrial networks, QKD services are provided by QKD repeaters/routers co-located with backbone nodes in passive optical networks preparing reserved quantum keys with prepare-and-measure QKD protocols. In non-terrestrial networks, QKD services are provided by QKD trusted relays, such as satellites and drones, preparing on-demand quantum keys through free space flexibly with entanglement-based QKD protocols. Moreover, we formulate the QKD resource placement optimization problem for the proposed system as a stochastic programming model. Numerical results show that the proposed system can successfully achieve the goal while satisfying all uncertain requirements and other constraints.
	\end{abstract}

	
	%%%%%%%%%%%%%%%%%%%%%%%%%%%%%%%%%%%%
	\section{Introduction}
	
\begin{figure*}[!]
		% 	\vspace{-0.18cm}
		\centering
		\includegraphics[width=0.75\linewidth]{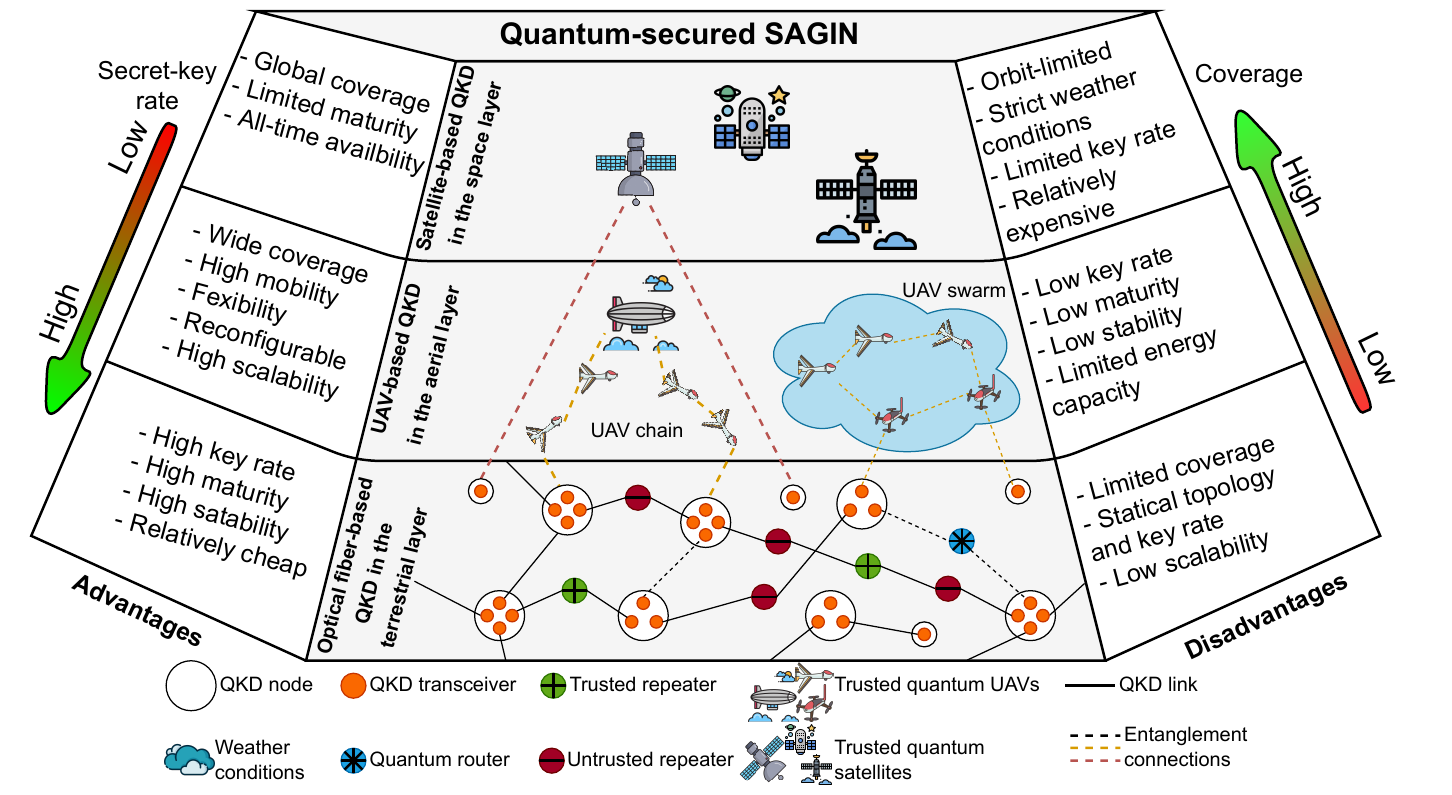}
		\caption{The concept of QKD over SAGIN in quantum-secured SAGIN.}
		\label{system}
	\end{figure*}

	% 第一段 from SAGIN to quantum-secured SAGIN
	% 主语一直是SAGIN
	% 1. SAGIN 很好
	% 2. 但是SAGIN中的通信安全需要被保护
	% 3. 有三种保护安全的方法, 对称,非对称, 和量子密码学
	% 4. 前面两种在量子时代已经不安全了
	% 5. 量子密码学能通过QKD的方式提供安全
	% 5. 因此SAGIN需要用QKD over SAGIN保护它里面的通信安全, 因此就有了quantum-secured SAGIN
	The growing interest in the sixth generation (6G) wireless networks, particularly the comprehensive connectivity and trustworthiness offered by space-air-ground integrated networks (SAGIN)~\cite{cui2022space}, is pushing the current communication infrastructure to its limits. However, secure communications of SAGIN are under serious threats in the post-quantum era. Currently, symmetric-key cryptography and public-key cryptography are used to protect the confidentiality of sensitive data in SAGIN, which are no longer considered safe with the advent of quantum computers \cite{quantum2022wang}. For example, the integer factorization and discrete logarithm problems can be easily compromised by a quantum computer that is advanced in computational power and quantum algorithms (e.g., Shor's algorithm). Fortunately, data channels of SAGIN can be secured by quantum cryptography, which uses quantum key distribution (QKD) to provision secret cryptographic keys via quantum channels and key management channels \cite{huang2021starfl}. In the post-quantum era, communications in quantum-secured SAGIN are expected to achieve the information-theoretic security with the guarantees of the principles of quantum physics, as guaranteed by the quantum no-cloning theorem and the Heisenberg’s uncertainty principle \cite{mehic2020quantum}.
	In quantum-secured SAGIN, the concept of QKD over SAGIN typically allows two remote QKD nodes to exchange quantum bits (qubits) which encode classical bits on quantum states, such as photons, through optical fibers or free space \cite{mehic2020quantum}. As shown in Fig. \ref{system}, the optical fiber-based QKD is a mature option based on the wavelength-division multiplexing (WDM) technique to transmit qubits and bits in the same fiber with a low loss and high stability. However, since qubits are more vulnerable to propagation impairments and cannot be readily amplified, the optical fiber-based QKD has limited distance requirements as its secret-key rate decreases exponentially when the distance increases. Therefore, as an alternative option, QKD via free space \cite{liao2017satellite, yin2020entanglement, liu2021optical}, e.g., satellite-based QKD and unmanned aerial vehicle (UAV)-based QKD, is more effective to transmit qubits with advantages of wide coverage and high flexibility. As a result, the concept of QKD over SAGIN, which comprises optical fiber-based QKD from the terrestrial layer, satellite-based QKD from the space layer, and UAV-based QKD from the aerial layer, provides a feasible solution for quantum-secured SAGIN.
	Although quantum-secured SAGIN is feasible through the concept of QKD over SAGIN, an efficient service provisioning framework is still necessary to manage and control QKD resources for optimizing the provisioning cost. To this end, a universal framework is proposed to offer secret-key distribution services to QKD nodes with security requests, i.e., secure communication requests, which may contain different secret-key sizes, rates, and updating periods \cite{mehic2020quantum}. By optimizing the deployment of QKD resources, such as QKD nodes and links, the framework can cost-effectively provision QKD services. In detail, the framework provides optical fiber-based QKD services statically with a fixed and relatively low cost. However, only the stationary QKD services may not be sufficient to secure the transmission of confidential data in dynamic communication environments. Therefore, the uncertainty in secure communication environments necessitates the provisioning of satellite- and UAV-based QKD services, according to the real-time secure communication requests flexibly with high cost. % Through the stochastic programming of QKD resources, this flexible framework is expected to resolve the uncertainty in security requests and minimize QKD service provisioning costs.
	% 这里需要改, 把上面的提出一个框架改成, 设计这个框架存在什么样的挑战
	% guaranteed secret-key rates and
	% Moreover, effective QKD service provisioning schemes for managing and controlling the QKD over SAGIN including quantum nodes and their quantum resources are necessary. The resources of a quantum node may include quantum memory, quantum channels, qubits, established quantum connections, etc. Such management methods can be used to monitor network status of the QKD over SAGIN, diagnose and identify potential issues (e.g. quantum connections), and configure quantum nodes with new actions and/or policies (e.g. to perform a new entanglement swapping operation). However, in QKD over SAGIN, the optical fiber-based QKD is determined by the network topology inherited from the existing passive optical network. Therefore, the optical fiber-based QKD cannot provide flexible QKD services for bursty communication traffic and security requirements. To satisfy the demands of secure communications in SAGIN, satellite-based and UAV-based QKD services are provisioned during the on-demand phase. 
	
	% case study of Metaverse in quantum-secured SAGIN
	% 主语是Metaverse
	% 1. 为了验证上述框架的有效性, Metaverse被作为用例使用quantum-secured SAGIn 为Metaverse 加密应用提供安全通讯.
	% 2. 元宇宙应用中的通信存在着许多的不确定和动态因素.
	% 3. 比如说, 用户通过AR/VR在元宇宙的3D世界中用他们的avatars获得沉浸式的体验. 由于用户可以同时操纵多个avatars在元宇宙内活动, 所以元宇宙中avatars的数量不可以被准确预测.
	% 4. 由于真实世界的环境的重要性不断变化, 元宇宙中的数字孪生的传输速度和更新频率也会有所不同.
	% 5. 元宇宙中AI模型的数量, 也会根据 AI 数据 和 计算能力 (e.g., number of workers in FL)的多少而存在变化.
	% 实验结果现实虽然为在quantum-secured SAGIn 为元宇宙应用提供QKD服务是一个复杂的问题, 但是所提出的框架仍然能最小化所需QKD服务提供的cost.
	To verify the effectiveness of the aforementioned concept and framework, the Metaverse~\cite{xu2022full} is used as a use case where users and Metaverse applications transmit data via quantum-secured SAGIN. The data generated and transmitted by the Metaverse applications, such as extended reality (XR), digital twin, and artificial intelligence (AI), are often communication-heavy, high-frequent, and asynchronous. For example, Metaverse users receive immersive experiences with their avatars in the 3D worlds of the Metaverse through XR, including augmented reality (AR), virtual reality (VR), and mixed reality (MR). Since each user can control multiple avatars in the Metaverse simultaneously, the number of avatars in the Metaverse cannot be accurately predicted by the QKD service provisioning framework. Second, due to the continuously varying importance of the real-world environment, the transmission speed and update frequency of the digital twins in the Metaverse also fluctuate. Third, the number of AI models in the Metaverse also varies depending on the amount of AI data and computational power (e.g., the number of workers in federated learning \cite{huang2021starfl}). Through the stochastic programming of QKD services, the proposed universal framework is expected to resolve the uncertainty issues in security requests and to minimize the provisioning cost. The experimental results demonstrate that although providing QKD services for Metaverse applications in quantum-secured SAGIN is an intricate problem, the proposed stochastic programming model can still optimize the required cost of QKD service provisioning.
	% 这里需要改, 首先我们要描述元宇宙的应用, 说明元宇宙的应用刚好存在这么多挑战的需求. 然后我们提出我们随机规划的方法, 为了应对元宇宙中对安全的随机, 我们就能够优化QKD服务提供, 并最小化提供的cost.
	%To address these issues, we propose a two-stage QKD service management scheme to minimized the cost of achieving quantum-secured communication over SAGIN. In reservation phase, the requesters of QKD services first reserve optical-fiber QKD resources in ground networks. Fiber-based networks can provide a large secret-key rates, but the resilience of their services is limited. Moreover, many QKD nodes in SAGIN that cannot be connected to optical fiber networks. Therefore, in the utilization and on-demand phase, the QKD requesters report their on-demand QKD demands to satellites and UAVs for remaining QKD services. Free-space QKD links can provide flexible QKD services with limited key materials, therefore can act as the supplement of the required secret-key rates in the on-demand phase.
	The main contributions of this article are summarized as follows:
	
	\begin{itemize}
		\item We highlight the feasibility of the quantum-secured SAGIN through the concept of QKD that we present. To the best of our knowledge, this is the first work to propose a quantum-secured SAGIN where the QKD services can holistically secure communications between space, air, and ground nodes using quantum cryptography.
		\item We propose a universal service provisioning framework to realize the concept of QKD over SAGIN. With uncertain communications and security requirements in quantum-secured SAGIN being taken into account, the proposed framework can provide flexible QKD services while minimizing the provisioning cost in quantum-secured SAGIN.
		% \item We propose a space-air-ground collaborative quantum-secured communication architecture that provides flexible QKD service provisioning for the uncertain and dynamic communications in SAGIN. In this architecture, uncertain communication and security requirements in SAGIN can be satisfied with limited QKD resources.
		\item As a case study, we are pioneering in examining the effectiveness of quantum-secured SAGIN and its service provisioning framework in supporting Metaverse applications. By considering the uncertain factors in the communications of Metaverse applications, the proposed framework can achieve an optimal provisioning solution for QKD services. The experimental results show that the proposed framework outperforms the existing baseline methods while addressing the uncertainty of communication environments with the varying number of avatars, digital twins, and AI models.
		% \item We formulate the QKD service provisioning problem in SAGIN as a stochatic integer programming to minimize the deployment cost pf QKD service provisioning. The proposed management scheme can improve the efficiency of decentralized federated learning while guaranteeing the security of model transmission, using secure federated learning as an example. The experimental results are realistic that the proposed scheme can secure the communication with high quality of service.
	\end{itemize}
	\section{The Concept of QKD over SAGIN and the Quantum-secured SAGIN} % 这个标题太难取了
	% 在这一章中， 我们主要讲述： QKD的基本概念(QKD是怎么进行密钥分发的)， 然后描述QKD over SAGIN的概念, 最后介绍QKD如何为Quantum-secured SAGIN建立安全通信
	\subsection{The Basic Concept of QKD}
	
	QKD has recently been introduced as a provable secure key distribution solution that can address many potential attacks, such as eavesdropping attacks, in traditional key distribution schemes \cite{cao2022evolution}. In practice, QKD is often adopted in the setup phase of secure communications between two nodes connected with QKD logical links for transmitting confidential data. During the setup of secure communications between two QKD nodes, they first need to agree on a global secret cryptographic key (i.e., a sequence of bits) via QKD links, where any eavesdropping or hijacking behaviors during key agreement process can be perceived due to the quantum no-cloning theorem \cite{quantum2022wang}. Therefore, in QKD logical links, their raw local secret keys are encoded as certain quantum states of photons and transmitted in quantum channels while the exchanged classical key information is verified and processed in the key management channels.
	
	\subsection{QKD over SAGIN}
	
	% Compared to classical signals, qubits are much more vulnerable to propagation impairments such as the scattering and loss over optical fibers as well as the atmospheric turbulence encountered by free-space optical links. Unfortunately they cannot be readily amplified, because amplifying the quantum signals would require measuring and cloning the quantum states, which is contrary to the quantum no-cloning theorem~\cite{buvzek1996quantum}. However, the fiber itself has losses and can lead to degradation of entanglement quality; and near-ground free space communication is affected by weather, obstacles and ground curvature; therefore, wide-area optical quantum transmission turns to satellite transit.
	
	\begin{table*}[!]
		\centering
		\caption{Summary of different QKD services in quantum-secured SAGIN}
		\begin{tabular}{|m{.1\textwidth}<{\centering}|m{.12\textwidth}<{\centering}|m{.2\textwidth}<{\centering}|m{.38\textwidth}<{\centering}|m{.08\textwidth}<{\centering}|}
			\hline
			QKD types                   & Transmission media       & QKD protocols                     & QKD resources     & Distance (km)                                                   \\ \hline
			Optical fiber-based QKD     & Optical fibers            & Prepare-and-measure/ Entanglement-based QKD protocols & Quantum transceivers, trusted/untrusted repeaters, quantum routers, quantum memory, QKD links      &$\sim$ 100               \\ \hline
			Satellite-based QKD & Free space & Entanglement-based QKD protocols  & Trusted quantum satellites, quantum memory, QKD links &$\sim$ 1000\\ \hline
			UAV-based QKD & Free space & Entanglement-based QKD protocols  & Trusted quantum UAVs, quantum memory, mobile QKD links & $\sim$ 1\\ \hline
		\end{tabular}
	\label{table:service}
	\end{table*}
	
	% 这张表, 加上distance 和 secret key rate
	% what is the structure of these three subsection, when we are introducing QKD services. We should high light that they are services!
	\subsubsection{Optical fiber-based QKD services}
	Existing conventional passive optical networks (PON) provide the maturest implementation scenario for fiber-based QKD in the terrestrial layer. With the WDM technique, qubits and bits can be transmitted together in the same optical fiber with low loss and high stability. As the optical fiber-based QKD is established upon the existing PON, its components are co-located with the backbone nodes in PON. There are three types of QKD nodes in optical-based QKD, i.e., quantum transmitters/receivers (transceivers), trusted/untrusted repeaters, and quantum routers. Firstly, each quantum transceiver is a pair of QKD equipment, including a quantum transmitter and a quantum receiver, that can convert bits of local secret keys into qubits and convert qubits into bits of global secret keys, respectively, via prepare-and-measure protocols, e.g., Bennett-Brassard-1984 (BB84), Grosshans-Grangier-2002 (GG02) \cite{cao2022evolution}. Then, the received secure keys are saved by their respective local key managers. Extending from the above point-to-point scenario, optical fiber-based QKD with trusted repeaters can distribute secret keys in a hop-by-hop manner to improve the key distribution distance. Furthermore, in the measurement-device-independent (MDI)-QKD protocol \cite{lo2012measurement}, untrusted repeaters typically have better security than other prepare-and-measure protocols requiring only trusted repeaters, which can remove the security infrastructure at the measurement side. Therefore, the untrusted repeater can even be controlled by an eavesdropper without compromising the safety of QKD. Therefore, MDI-QKD protocols are capable of improving the secure distance and efficiency of QKD considerably. Different from prepare-and-measure protocols, entanglement-based protocols, such as Bennett-Brassard-Mermin-1992 (BBM92) \cite{yin2020entanglement}, implement quantum routers with quantum memory to distribute secret keys via quantum entanglement of photons over different quantum channels. Between two adjacent quantum routers, each external link is established where one of the qubits in a Bell pair should be sent to the other quantum router of this external link, and hence one unit of quantum memory on each router needs to be occupied \cite{zhao2021redundant}. Then, quantum routers perform entanglement swapping to measure the qubits and route the result to the destination router via the internal link of an entanglement connection. For the optical fiber-based QKD services in the terrestrial layer, quantum transceivers and QKD links of trusted/untrusted repeaters as well as quantum memory of quantum routers are collectively regarded as QKD resources in provisioning QKD services over SAGIN.

	\subsubsection{Satellite-based QKD services}
	Different from the optical fiber-based QKD in the terrestrial layer, the satellite-based QKD in the space layer can provision QKD services to ground QKD nodes from low earth orbit (LEO) satellites via free space \cite{liu2020drone}. Due to scattering and absorption of polarized photons and other factors, the requirements of trusted/untrusted repeaters to extend the coverage of optical fiber-based QKD place an obstacle to developing metropolitan-level QKD networks. Fortunately, satellite-based QKD in the space layer is a promising scheme to distribute secret keys to ground QKD nodes via free space with less atmospheric attenuation. China sent its first quantum satellite, called Micius, into space in 2016 and completed an experimental satellite-to-ground QKD in good weather and night conditions. In June 2020, Micius experimentally demonstrated that satellite-based QKD could achieve a secret-key rate of 0.12 bits per second between two ground QKD nodes separated by 1120 km \cite{yin2020entanglement}. In satellite-based QKD, satellites with are regarded as fully trusted quantum relays to establish entanglement connections between a pair of ground QKD nodes with polarization beam splitters to extend the coverage of optical fiber-based QKD.

	\subsubsection{UAV-based QKD services}
	
	Since satellites are orbit-limited and weather-constraint that are not able to satisfy all the security requests on the ground, trusted UAVs carrying quantum equipment can provide a reconfigurable solution to serve ground QKD nodes regardless of their weather conditions. By transmitting quantum information as ``flying quantum bits", UAVs in the aerial layer can secure communications in SAGIN with a high degree of mobility and flexibility \cite{liu2020drone}. Specifically, two UAVs can achieve mobile entanglement distribution via free-space quantum channels and key management channels, with one distributing entangled photons and the other acting as a relay node. However, beam-diffraction, which depends on the beam aperture and the link distance, limits the distance of the UAV-based QKD via free space. To address these issues, multiple UAVs can form swarms or chains, to distribute the quantum entanglement collaboratively for lossless propagation within the Rayleigh length limit. The experimental results in \cite{liu2021optical} show that UAV-based QKD can achieve high-performance quantum entanglement between two ground QKD nodes with polarization beam splitters at a distance of about 1 km. Due to the inexpensive and flexible nature of UAV swarms and chains, UAV-based QKD can provide reconfigurable QKD services in an economical manner.
	
	\subsection{Establishing Secure Communications in Quantum-secured SAGIN}
	In quantum-secured SAGIN, a pair of two QKD nodes can establish secure communications via QKD links for transmitting confidential data. As shown in Table \ref{table:service}, If QKD nodes are directly or indirectly connected via optical fibers, they can continuously receive secret keys from optical fiber-based QKD services. Otherwise, QKD nodes can require the QKD services via free space from quantum satellites or quantum UAVs. On the one hand, satellite-based QKD services can distribute quantum entanglement in global coverage but are limited by the orbits of satellites and the weather conditions of ground nodes. On the other hand, UAV-based QKD can provision reconfigurable QKD services regardless of the location and weather condition of ground nodes. Through the concept of QKD over SAGIN, the quantum-secured SAGIN is feasible with unparalleled mobility, flexibility, and reconfigured by provisioning the QKD services via optical fibers, satellites, and UAVs synergistically.
	
	\section{The Universal QKD Service Provisioning Framework in Quantum-secured SAGIN}
	
	\begin{figure}[!]
		% 	\vspace{-0.18cm}
		\centering
		\includegraphics[width=0.8\linewidth]{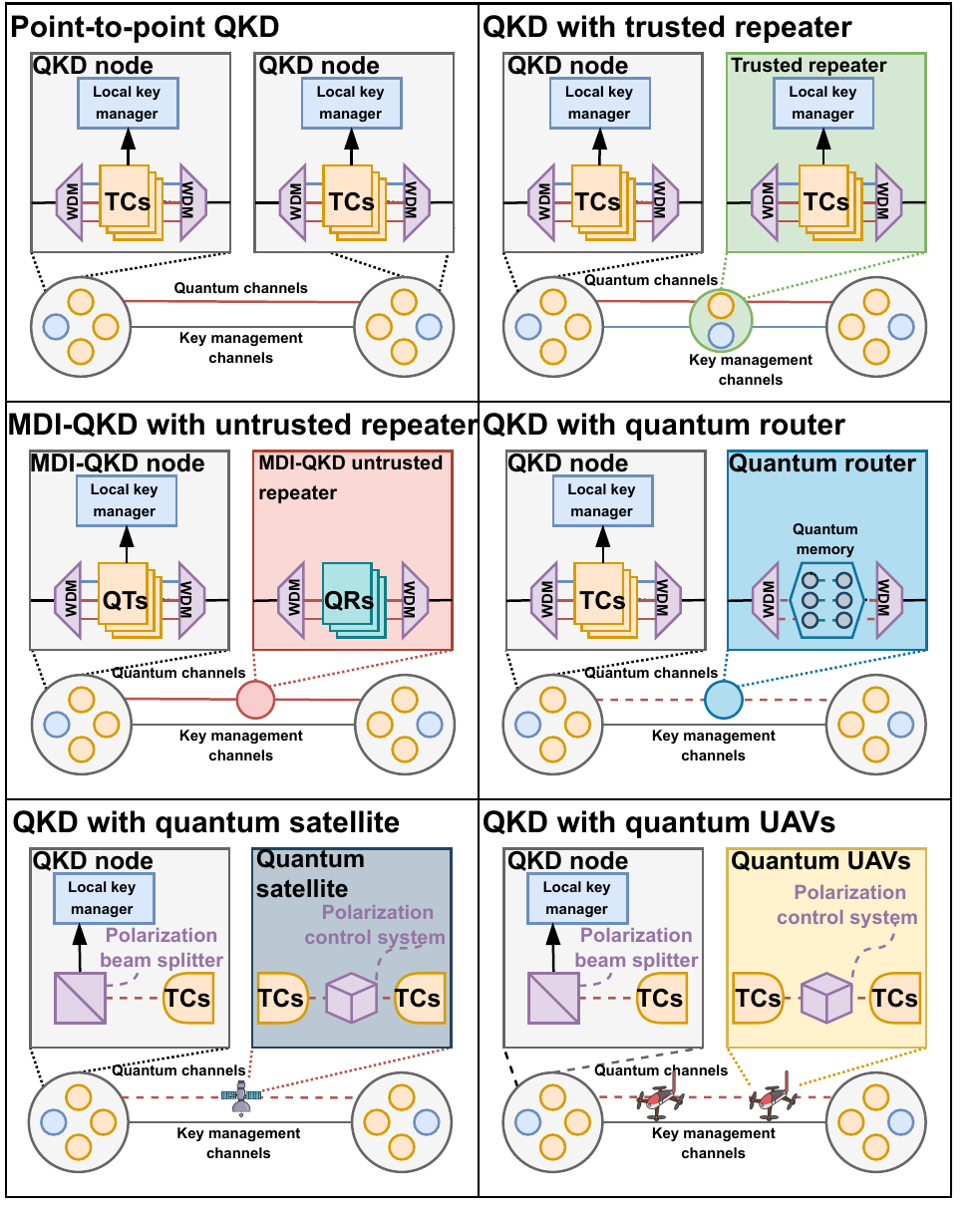}
		\caption{Different types of QKD resources in quantum-secured SAGIN. TCs, quantum transceivers; QTs, MDI-QKD transmitters; QRs, MDI-QKD receivers.}
		\label{QKDresource}
	\end{figure}
	
	% For large-scale QKD, QKD over SAGINs~\cite{zhang2018large} (QKDN) are required, which can be regarded as a subset of a Quantum Internet. A QKDN may consist of a QKD application layer, a QKD over SAGIN layer, and a QKD link layer. One or multiple trusted QKD relays may exist between the source quantum node A and the destination quantum node B, which are connected by a QKDN. Alternatively, a QKDN may rely on entanglement distribution and entanglement-based QKD protocols; as a result, quantum-repeaters/routers instead of trusted QKD relays are needed for large-scale QKD.

	\subsection{Provisioning Static QKD Services via Optical Fiber-based QKD}
	
	In the terrestrial layer of the quantum-secured SAGIN, optical fiber-based QKD is able to provision static QKD services by allocating QKD resources in terms of QKD nodes, e.g., quantum transceivers, trusted/untrusted repeaters, quantum routers, and QKD links, as shown in Fig. \ref{QKDresource}. In quantum-secured SAGIN, cryptographic applications initiate requests for QKD services to the QKD global manager. After receiving these QKD service requests, the QKD global manager sends instructions to the optical fiber-based QKD controller. If the secret keys in the local key managers of the requesting QKD nodes are sufficient to satisfy these requests, then the optical fiber-based QKD controller configures the QKD nodes and lets them provide the secret key to the cryptographic applications. However, since the coverage of optical fiber-based QKD services is limited and fixed, the optical fiber-based QKD services are not able to provide sufficient secret keys to well handle the uncertainty and dynamics in secure communications of QKD nodes.
	% To provide secure key distribution services to the nodes in the quantum network. QKD nodes in the terrestrial-networks first need to deploy QKD devices such as quantum transceivers and relays. In this way, QKD can generate quantum keys in a certain secret-key rate depending on the number of reserved QKD resources in terms of QKD devices and QKD links.
	
	\subsection{Provisioning Dynamic QKD Services via Satellite- and UAV-based QKD}
	
	Fortunately, satellite- and UAV-based QKD services can supplement the required secret keys for QKD nodes to resolve the uncertain security requests in their communications with high coverage and flexibility. On the one hand, when the secret keys are not enough in the terrestrial layer, the satellite-based QKD controller can configure LEO satellites to distribute the required QKD services via satellite-based QKD in the space layer. Although satellite-based QKD services are real-time and have global coverage, they are orbit-limited and constrained by the weather conditions of the ground nodes. Therefore, on the other hand, UAV-based QKD in the aerial layer may be a practical alternative to provision QKD services at various times, locations, and weather conditions. According to the instructions from the QKD global manager during the data transmission phase, the UAV-based QKD controller can deploy UAV swarms or chains to provide QKD services in mobile quantum networks \cite{liu2020drone}.
	
	% Although data traffic is usually heterogeneous and uncertain in the Internet, the reserved QKD services may not enough to guarantee the security demands of network nodes. As a result, network nodes are required to allocate extra on-demand QKD services from non-terrestrial QKD over SAGINs. Also, some network nodes that are not connected to the PON via fiber need to request on-demand QKD services from non-terrestrial QKD over SAGINs to meet their secure communication requirements. Finally, although non-terrestrial QKD over SAGINs can freely provide quantum keys to network nodes, there is an overall maximum rate of quantum keys that they can provide. At the same time, non-terrestrial QKD over SAGINs have very stringent requirements on the outside conditions of QKD nodes. In general, satellite-based and UAV-based QKDs need to be performed at night. In addition, the success of quantum key distribution depends on the weather conditions of the QKD node at the receiving end.
	
	\subsection{Challenges of Provisioning QKD Services in Quantum-secured SAGIN}
	
	\subsubsection{Efficient Routing for QKD Services}
	% First, the distance of the qubit transmission is limited
	% Second, the reliance of trusted relays
	% Quantum-secured SAGIN is an overlay networks, where the data transmission in the upper layer data networks is limited by the bottom layer QKD services.
	To provide efficient QKD services in quantum-secured SAGIN, many practical challenges are required to be solved. First and foremost is the problem of point-to-point routing in optical fiber-based QKD in the terrestrial layer. Unlike routing in traditional backbone networks, optical fiber-based QKD requires nodes on the path of secure communications to be fully active and have enough secret keys. As shown in Fig. \ref{system}, there are two relaying nodes for the current optical fiber-based QKD, i.e., trusted/untrusted repeaters and quantum routers. In optical fiber-based QKD services with trusted/untrusted repeaters, the nodes are connected with the intermediate of trusted/untrusted repeaters and the secret keys as well as key management information through QKD links and the encrypted message are transmitted through public data channels. Each QKD link is always connected in a point-to-point fashion. At the same time, there are a variable number of intermediate nodes in public data channels, regardless of whether they are trusted. In optical fiber-based QKD services with quantum routers, each quantum router has a number of quantum memories for storing a finite number of quantum bits and a certain number of quantum entangled links per variable. Unlike the store-and-forward mechanism of trusted repeaters, the motivation of the implementation of quantum routers is to apply the quantum entanglement of photons to transmit quantum signals over different quantum channels. In other words, multiple quantum memory are connected together and the measurement of one of the quantum memory can change the quantum states of the other quantum memory. Even when a considerable distance separates the photons, they still establish a joint quantum system. For optical fiber-based QKD services, each distribution of the secret key information consumes a certain amount of quantum resources of QKD nodes. Since the quantum resources in each node are limited, the secret keys of each QKD node are also limited. The purpose of efficient routing for QKD services is to minimize the secret key consumption, i.e., the number of hops transmitted, while satisfying the secure communication requests \cite{mehic2019novel}.
	
	\subsubsection{Uncertainty and Dynamics of Secure Communication Environments in Quantum-secured SAGIN} 
	% 承接上面的overlay networks, 在reservation阶段, QKD
	Quantum-secured SAGIN still needs an effective dynamic resource management scheme to enhance the efficiency in provisioning QKD services. The communication security in quantum-secured SAGIN is guaranteed by secret keys. The encryption operation of QKD nodes cannot be implemented without the secret keys in their local key managers, and thus the encrypted communication link is temporarily invalid. Although one-time pad encryption can provide information-theoretically secure (ITS) communication, due to the limited secret key provided by optical fiber-based QKD services, the requirement for the one-time pad to have an encryption key of equal length to the plaintext is difficult to achieve \cite{cao2022evolution}. Therefore, the current solution of quantum-secured SAGIN generally updates the secret keys periodically to ensure a certain level of secure communications. Nevertheless, before the transmission of confidential data, the data rate is unpredictable and QKD services provisioned by optical fiber-QKD are fixed and undersubscribed. Therefore, allocating satellite- and UAV-based QKD services to compensate for the secret key shortage is challenging to resolve the uncertainty and dynamics of secure communication environments.

	\section{Case Study: Metaverse in Quantum-secured SAGIN}
	
	\begin{figure*}[!]
		% 	\vspace{-0.18cm}
		\centering
		\includegraphics[width=0.75\linewidth]{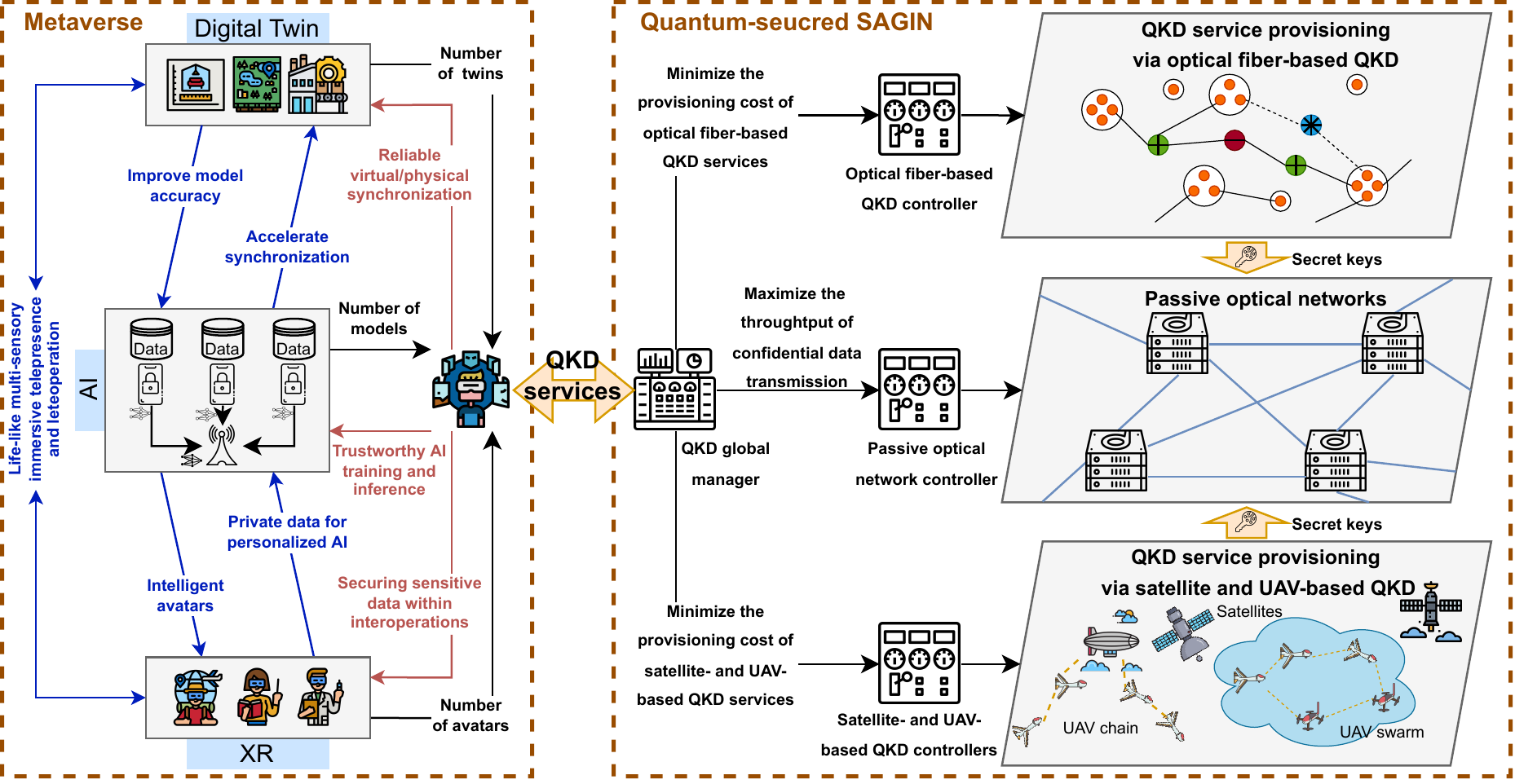}
		\caption{The universal QKD service provisioning framework with two-stage stochastic programming for Metaverse application in quantum-secured SAGIN.}
		\label{framework}
	\end{figure*}

	As illustrated in Fig. \ref{framework}, in this case study of Metaverse in quantum-secured SAGIN, we first provide an overview of Metaverse and highlight the requirements for secure interoperations of Metaverse applications. To provide QKD services for Metaverse applications, we develop the system model, provisioning options, and serving phases for QKD over SAGIN. Finally, we describe the uncertainty of Metaverse applications and propose an optimization solution to provide QKD services cost-effectively.
	
	\subsection {Metaverse Applications}
	The Metaverse, considered as the successor to the mobile Internet, is a set of interoperable 3D worlds where people can telepresent themselves to work, play, and socialize \cite{wang2022survey}. In the Metaverse, applications/users provide seamless interoperations with other applications/users by constantly communicating with each other. As shown in Fig. \ref{framework}, interoperations between XR and digital twin are required to provide life-like, multi-sensory, immersive telepresence and teleoperation between users and the real world. AI can be used to accelerate physical/virtual synchronization in the digital twin, and AI can then use the sensory data to improve the accuracy of AI models. In addition, Metaverse users enter the Metaverse in the form of avatars and interact with the Metaverse's intelligent avatars while using their private data to create personalized AI models. However, the embodied experiences provided by Metaverse applications make Metaverse users more vulnerable to threats of unsafe interoperability, such as unreliable virtual/physical synchronization, untrustworthy AI training and inference, and sensitive data leakage. Especially, in the post-quantum era, quantum computing is believed to be the cornerstone of the Metaverse applications and can be used for XR rendering, digital twin scientific calculations, and AI model training. However, quantum computing can compromise the secure interoperations of Metaverse applications based on traditional symmetric-key cryptography and public-key cryptography. 
	Fortunately, QKD can provide information-theoretically secure interoperations for Metaverse applications. Moreover, QKD over SAGIN can secure the interoperations of Metaverse applications in a timely and global manner via quantum-secured SAGIN.

	\subsection{System model, Provisioning Options, and Serving Phases}
	
	To provision the QKD services to Metaverse applications, the system model of quantum-secured SAGIN includes data nodes, QKD nodes and edges \cite{cao2021hybrid}. Here, data nodes are co-located with the QKD nodes. Edges in quantum-secured SAGIN are QKD links, e.g., optical fibers and entanglement connections, depending on the equipment of QKD nodes. For optical fiber-based QKD, each QKD node consists of a local key manager, and one or more transceivers of MDI-QKD. Between a pair of QKD nodes linked with optical fiber, a QKD link with multiple trusted/untrusted repeaters can be deployed for global secret key generation. Each trusted repeater consists of two or more QKD transceivers, a local key manager, and security infrastructure, while each untrusted repeater consists of two or more QKD transceivers but with a higher security level. Between a pair of QKD nodes, quantum routers with a small amount of quantum memory can establish entanglement connections for entanglement swapping. For satellite- and UAV-based QKD, trusted quantum satellites and trusted quantum UAVs act as quantum relays.
	
	The QKD global manager can offer secure communications of Metaverse applications in quantum-secured SAGIN with two provisioning options, i.e., QKD via optical fiber and QKD via free space. In general, the option of QKD via optical fiber provides QKD services through optical fiber-based QKD, while the option of QKD via free space provides QKD-supported secure communication services through satellite- and UAV-based QKD. However, the QKD via free space is more dynamic and its use can be on a short-term on-demand basis. It is important to note here that these provisioning options are similar to that of other network and cloud services available commercially, i.e., reservation and on-demand subscription plans.
	
	QKD service provisioning in quantum-secured SAGIN is divided into three serving phases by the QKD global manager, i.e., QKD via optical fibers, data transmission, and QKD via free space. First, in the phase of optical fiber-based QKD phase, the QKD global manager follows the option of QKD via optical fiber to instruct the optical fiber-based QKD controller to deploy optical fiber-based QKD services. As this should be done on a long-term basis, this option could be done without knowing the secure communication requirements of Metaverse applications. Then, when the option of QKD via optical fiber is determined, quantum-secured SAGIN enters the data transmission phase. As the QKD via optical fiber is more stable and can be done on a long-term reservation basis, the secret keys provided by optical fiber-based QKD services are static. During the data transmission phase, if the security requirements of Metaverse applications incurs more secret keys which cannot be provided only by optical fiber-based QKD services, the Metaverse applications can request additional QKD services from the QKD global manager in the option of QKD via free space and then obtain more secret-key rate from the QKD via free space option. In this phase, the QKD global manager sends instructions to the satellite- and UAV-based QKD controllers to configure QKD services for Metaverse applications requiring extra QKD services. For the same secret-key rate, the provisioning cost of QKD via optical fibers is typically less than that of QKD via free space, e.g., due to the network operating costs. The QKD global manager aims to satisfy the security requests of Metaverse applications while minimizing the provisioning cost of QKD services.
	
	\subsection{Uncertainty of Metaverse Applications in Quantum-secured SAGIN}
	
	Under uncertainty of secure interoperations of Metaverse applications, the secret key rate of required QKD services by Metaverse applications is not precisely known when the option of QKD via optical fibers is made. Moreover, since the data transmission rates in the Metaverse fluctuate over time, the Metaverse applications cannot predict the required QKD services precisely. For example, due to the fact that AI data quantity and quality generated in Metaverse applications are varying over time, the frequency and size of updates to AI models are uncertain. Second, the number of digital twins is unpredictable and the synchronizing sizes and frequencies are varying according to the changing importance of the real-world environment. Third, the number of avatars in the Metaverse is dynamic, as each user can have multiple avatars simultaneously in different virtual 3D worlds. In general, the probability distribution of uncertainty can be partially fitted by statistical processes and machine learning.
	
	\subsection{The Proposed Optimization Solution Approach}
	
	In the deterministic programming model for QKD service provisioning in quantum-secured SAGIN, the required secret keys in QKD services of Metaverse applications are assumed to be exactly known by the QKD global manager. However, due to the aforementioned challenges in the QKD service provisioning framework and uncertain factors of secure communication environments in Metaverse applications, the QKD service requests for secure communications cannot be predicted precisely and fulfilled perfectly. Thus, the stochastic programming with two-stage QKD service provisioning is proposed. The first stage defined the number of optical fiber-based QKD services provisioned in the terrestrial layer, while the second stage defines the number of satellite- and UAV-based QKD services deployed in the space layer and the aerial layer. The stochastic programming model can be formulated as the minimization of provisioning cost including two parts, where the first part is the provisioning cost of optical fiber-based QKD services according to the option of QKD via optical fibers and the second part is the expected provisioning cost of the option of QKD via free space under the set of possible realized data transmission requests of Metaverse applications. The constraints of the stochastic programming model for QKD over SAGIN are flow conservation, the required number of wavelengths, transmission path uniqueness, wavelength continuity, wavelength capacity, and wavelength uniqueness \cite{cao2021hybrid} under the uncertainty of secure communication environments in Metaverse applications.
	
	\subsection{Experimental Results}
	
	\begin{figure}[t]
		\centering
		\subfigure[Cost structure.]{
			\begin{minipage}[t]{0.48\linewidth}
				\includegraphics[width=1\linewidth]{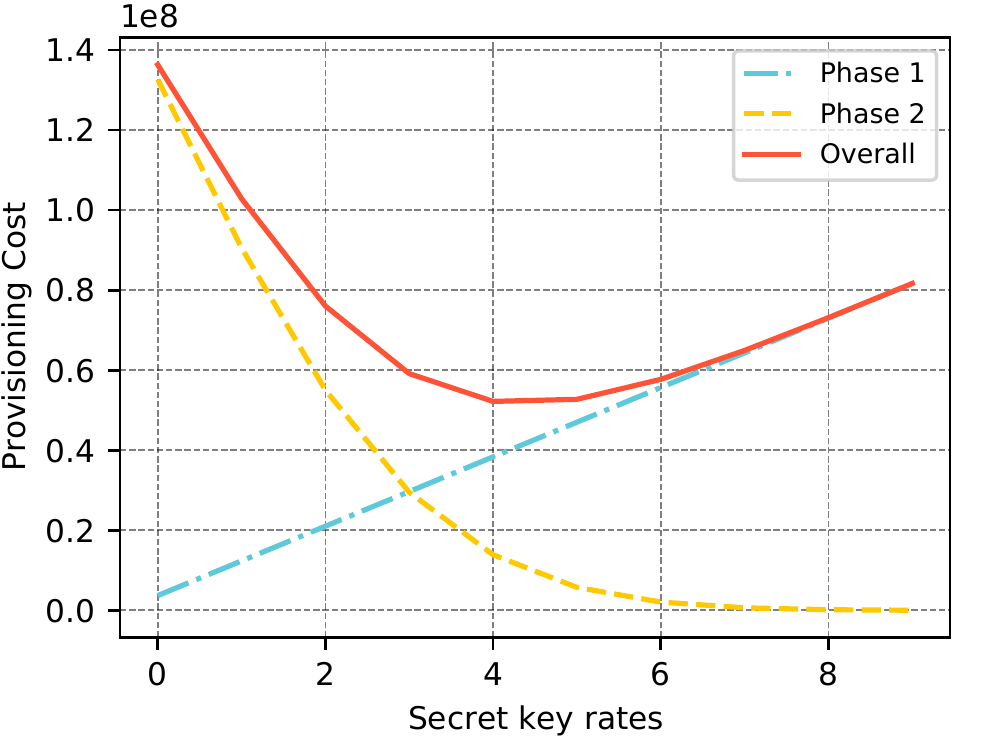}
				%\caption{fig1}
			\end{minipage}%
		}%
		\subfigure[Performance comparison.]{
			\begin{minipage}[t]{0.48\linewidth}
				
				\includegraphics[width=1\linewidth]{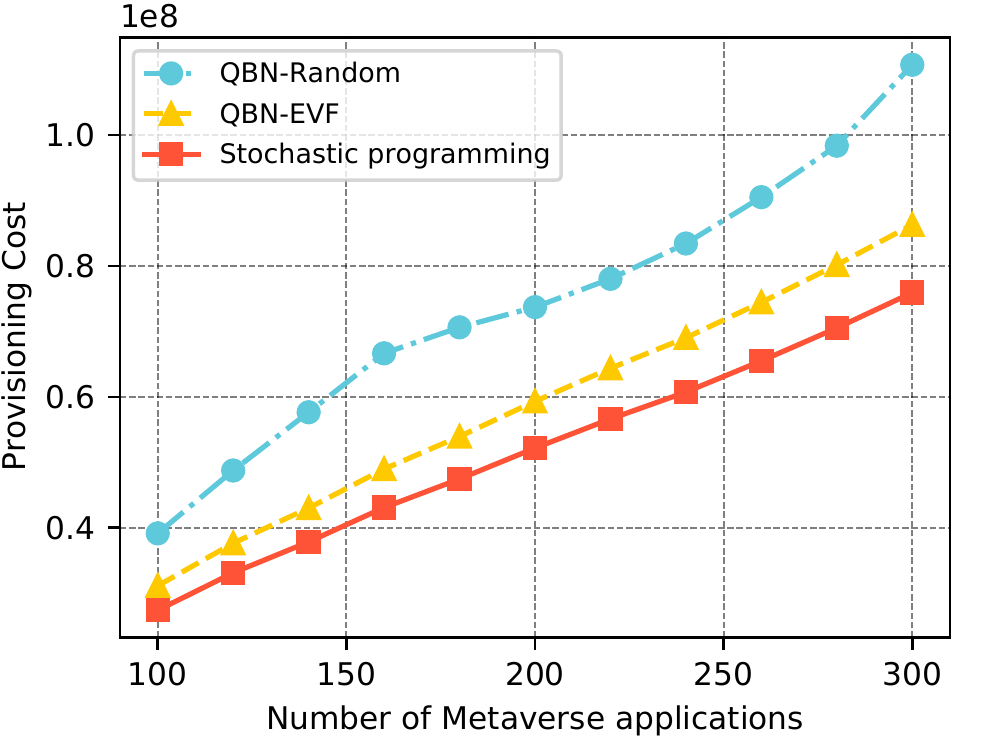}
				%\caption{fig2}
			\end{minipage}%
		}%
		% 	\subfigure[USNET.]{
			% 		\begin{minipage}[t]{0.33\linewidth}
				
				% 			\includegraphics[width=1\linewidth]{magazine_USNET_ondemand.pdf}
				% 			%\caption{fig2}
				% 		\end{minipage}%
			% 	}%
		\caption{The cost structure and performance comparison of stochastic programming for secure communications of Metaverse applications in quantum-secured SAGIN.}
		\label{experiment}
	\end{figure}
	
	The experiments are performed on the USNET topology with 24 nodes \cite{cao2022evolution}. The distance between each pair of QKD nodes in optical fiber-based QKD is set to 80 km and the distance between each pair of UAVs in UAV-based QKD is set to 1 km. The required secret keys of Metaverse applications are set to the equivalent level. To simplify the evaluation, the uncertainty is reformulated as all the Metaverse applications have the equivalent requested size of secret keys that is determined by the amount of scenarios. For each scenario, the probability distribution of secret keys requirements follows a Poisson process with an average of the number of scenarios (set to ten) divided by three. The default number of Metaverse applications is set to 200. Finally, the reservation cost values, with a unit denoting a unit of normalized monetary, of QKD transmitters, QKD receivers, local key manager, security infrastructure, MUX/DEMUX components, and optical fiber, are 1500, 2250, 1200, 150, 300, and 1, respectively. The on-demand cost values, with a unit representing a normalized monetary unit, of QKD transmitters, QKD receivers, securet-key buffers, security infrastructure, UAVs, and satellites are 6000, 9000, 3000, 500, 20, and 20000, respectively. QKD devices are relatively expensive, however, a small secret key can accomplish the encryption of a large amount of confidential data e.g., a 256-bit secret key can be used to encrypt 64 GiB data in AES-256-GSM. Furthermore, new quantum infrastructures are required to be deployed for QKD service provisioning which will be the fundamental deployment of the upcoming quantum Internet \cite{quantum2022wang}.
	
	Based on the aforementioned experimental settings, the cost structure of stochastic programming of QKD services in quantum-secured SAGIN is first studied. As illustrated in Fig. \ref{experiment}(a), the first phase, the second phase, and the overall provisioning costs vary as the number of reserved QKD services increases. As expected, the first phase provisioning cost linearly increases as the number of reserved QKD services rises. Nevertheless, during the data transmission, the second phase cost decreases as the number of reserved QKD services increases, so that less on-demand compensation is required for Metaverse applications. In this way, the optimized provisioning plan can be determined by the minimization of overall cost, e.g., when the size of secret keys provided by optical-based QKD services equals to four as illustrated in Fig. \ref{experiment}(a). Through cost analysis in QKD service provisioning framework, the optimal provisioning plan can not easily be obtained due to the uncertainty in the secure interoperations of Metaverse applications. For example, the optimal QKD service provisioning solution is not the place where the cost curves of the two phases of service are intersected. Therefore, a stochastic programming model for QKD over SAGIN is essential to achieve the minimized service provisioning cost.
	
	In the experiments, two baseline algorithms, i.e., the quantum backbone network (QBN) model with expected value formulation (QBN-EVF) provisioning and the QBN with random (QBN-Random) provisioning, developed in~\cite{cao2022evolution}, are used for the compassion of the stochastic programming model. In QBN-EVF, the QKD services provisioned in the first phase are fixed by the expectation of the requested security requests, which is an approximate model. Similarly, in the QBN-Random model, the secret keys provisioned by the optical fiber-based QKD services are generated uniformly from zero to the number of scenarios, indicating a random model. The provisioning costs of QBN-Random, QBN-EVF, and stochastic programming for different number of security requests in Metaverse applications are shown in Fig. \ref{experiment}(b). We can observe that as the number of Metaverse applications increases, the provisioning costs of all three models increase accordingly. Moreover, the performance difference in provisioning cost among the three models is also larger as the number of Metaverse applications increases. In particular, the provisioning cost of QBN-EVF is slightly higher than that of the stochastic programming model, while the provisioning cost of QBN-Random is 50\% higher.
	
	\section{Conclusion}
	
	In this article, we have highlighted the feasibility of quantum-secured SAGIN through the conceptualization of QKD over SAGIN. We have presented the concept of QKD over SAGIN, where optical fiber-based QKD, satellite-based QKD, and UAV-based QKD services are provisioned collaboratively to secure communications in SAGIN. To realize the concept of QKD over SAGIN, we have proposed a universal QKD service provisioning framework which is expected to cater flexible and cost-efficient QKD services in quantum-secured SAGIN. Finally, we have examined the proposed concept and framework by using Metaverse applications as a case study. The experimental results have demonstrated that proposed methods can minimize the provisioning cost under uncertain secure communication environments.
	
% There are several possible directions that are worth being studied: i) intelligent algorithms for efficient QKD service provisioning in optical fiber-based QKD. ii) social-aware relaying schemes for QKD service provisioning with hybrid trusted/untrusted repeaters. iii) economic designs for the universal QKD service provisioning framework.
	
	\bibliographystyle{ieeetr}
	\bibliography{myreference}
\end{document}